# Layer-number-dependent spin Hall effects in transition metal monocarbides $M_2$C ($M$=V, Nb, Ta)

Xi Zuo[1], Yulin Feng[1], Na Liu[1], Bing Huang[2], Desheng Liu[3*], Bin Cui[3*]

[1]College of Physics and Electronic Science, Hubei Normal University, Huangshi 435002, China.

[2]Beijing Computational Science Research Center, Beijing 100193, China.

[3]School of Physics, National Demonstration Center for Experimental Physics Education, Shandong University, Jinan 250100, China.

*emails: liuds@sdu.edu.cn; cuibin@sdu.edu.cn

## Abstract

The recent discovery of strong spin Hall effects (SHE) in 2D layered topological semimetals has attracted intensive attention due to its exotic electronic properties and potential applications in spintronic devices. In this paper, we systematically study the topological properties and intrinsic SHE of layered transition metal carbides $M_2$C ($M$=V, Nb, Ta). The results show that both bulk and monolayer $M_2$C have symmetry-protected nodal points (NPs) and lines (NLs) originating from the $d$ band crossing near the Fermi level ($E_F$). The inclusion of SOC breaks the degeneracy of NLs and NPs, contributing to large spin Hall conductivity (SHC) up to ~1100 and ~200 $(\hbar/e)(\Omega\ \text{cm})^{-1}$ for bulk and monolayer Ta$_2$C, respectively. Remarkably, we find that magnitude of SHC exhibits a significant enhancement by increasing the layer number. For eight-layer Ta$_2$C, the maximum value of SHC can reach up to ~600 $(\hbar/e)(\Omega\ \text{cm})^{-1}$, comparable to many reported 3D topological materials. Analysis of spin Berry curvature reveals that the large SHC originates from layer-number-dependent nodal line structure near the $E_F$, in which the repeated crossover between valence and conduction bands creates large amounts of NPs along the Γ-K route. Our findings not only provide a new platform for experimental research of low-dimensional SHE, but also suggest an effective way of realizing giant SHE by controlling layer thickness.

# I. INTRODUCTION

Spin Hall effects (SHE), a relativistic phenomenon that a transverse spin current generated by an applied electric field in the absence of a magnetic field, has become an important topic in recent years [1,2]. Two key factors, intrinsic spin Hall conductivity (SHC) and spin Hall angle (SHA), influence the performance of SHE-based devices. The intrinsic SHC can be accurately calculated by integrating the spin Berry curvature (SBC) of the occupied bands in the Brillouin zone (BZ) [3]. And the SHA of a SHE system is the ratio of the SHC to the charge conductivity ($G_C$), which represents the charge-to-spin interconversion efficiency at room temperature [4]. Therefore, exploring SHE systems with large SHC and SHA is one of the main goals in this field.

With the rapid development of topological matters, strong SHE is observed in topological systems such as topological insulators (TIs) and topological semimetals (TSMs) [5–7]. Due to their spin-momentum-locked surface states, TIs are considered to be ideal systems for generating pure spin currents. However, various experiments show that the SHA of the $Bi_2Se_3$ family can vary from 0.01 to 425 due to the bulk doping problem induced by the hybridization between surface states and bulk states [5,8,9]. Recently, increasing attention has been focused on the intrinsic SHE in TSMs. For example, nodal point semimetals ZrSiTe [10], TaAs [11], and LaAlGe [12] have been predicted to host large intrinsic SHC exceeding 500 $(\hbar/e)(\Omega\ cm)^{-1}$ due to large SBC around the spin-orbit coupling (SOC) induced small gapped Dirac or Weyl nodal points. On the other hand, large intrinsic SHEs have also been reported in nodal line semimetals. Different from nodal point semimetals, the gapped nodal lines can induce many band anticrossing points distributed with large and continuous SBC in the entire BZ, contributing to a remarkable SHC [13]. It is predicted that NLSMs such as $W_3Ta$ [14], $Ta_3As$ [15], $ThAl_2$ [16], and InBi [17] with multiple nodal lines around the Fermi level ($E_F$) exhibit a giant SHC exceeding 1000 $(\hbar/e)(\Omega\ cm)^{-1}$. To realize large intrinsic SHC and SHA as well as to design new devices for spin-charge current conversion, it is important to explore new TSMs and study the interplay between the SHC and the band topology.

Recently, the 2D transition metal dichalcogenides (TMDs) family are reported to exhibit large SHE due to their tunable SOC, $G_C$, and band topology [18–20]. For example, the Weyl semimetals MoTe$_2$ and WTe$_2$ are reported to own large SHA of 0.32 [21] and 0.17 [22], respectively. Spin-orbit torque in few-layers Dirac semimetal PtTe$_2$ [23] reveals that the SHC exhibits a monotonical increment with a maximum value of about 1000 $(\hbar/e)(\Omega\ \text{cm})^{-1}$ as the layer thickness increases from 3 nm to 20 nm. Correspondingly, first-principle calculations on bilayer PtSe$_2$[24] and tri-layer MoTe$_2$ [21] reveal that its intrinsic SHC is about 27 and 200 $(\hbar/e)(\Omega\ \text{cm})^{-1}$ at the $E_F$, respectively. This raises the question of whether we can find large intrinsic SHC in layered 2D materials comparable to familiar heavy metals or 3D topological materials. On the other hand, the influence of layer thickness on the SHE in 2D materials still calls for further theoretical investigation.

In this paper, we systematically investigate electronic structures and intrinsic SHE in 2D layered transition metal carbides $M_2$C ($M$=V, Nb, Ta, known as MXenes), which exhibit a fascinating combination of properties such as controllable minimum layer thickness, large electric conductivity, rich electronic structures, and can be easily synthesized in the laboratory [25–30]. Both bulk and monolayer $M_2$C exhibit large SHC owing to strong SOC and contributions of multiple nodal lines in the band structures. Taking Ta$_2$C as an example, we observe a significant enhancement of SHC by varying the thickness from one layer (1L) to eight layers (8L), which originates from a layer-number-dependent nodal line structure near the $E_F$. The rest of this article is organized as follows. In Sec. II A, we first report the calculated band structures, SHC, and SHA for bulk $M_2$C. In Sec. II B, we then present the calculated SHC ($\sigma_{xy}^z$) by varying the layer number from 1L up to 8L. To understand the origin of layer-number-dependent SHC, we also give the electronic band structure, nodal line structure, and $k$-resolved SBC at the $E_F$. Finally, the conclusion drawn from this work is summarized in Sec. III.

## II. RESULTS AND DISCUSSION

## A. The electronic band structures and SHE in bulk M2C

As shown in Fig. 1(a), the bulk $M_2C$ ($M$=V, Nb, Ta) crystallize in a trigonal crystal structure with the space group $P\bar{3}m1$ (164). The C atom locates at the corner sites of the unit cell, while two $M$ atoms locate at equivalent sites with the Wyckoff position of 2d (1/3, 2/3, 1/4). Each C atom is bonded to six equivalent $M$ atoms to form an edge-sharing octahedral, and each $M$ is bonded in a distorted T-shaped geometry to three equivalent C atoms. The optimized lattice parameters and bond lengths are summarized in Table S1 in the Supplemental Material (SM) [31], which agree well with previous experimental and theoretical values [29,30]. The BZ is shown in Fig. 1(b).

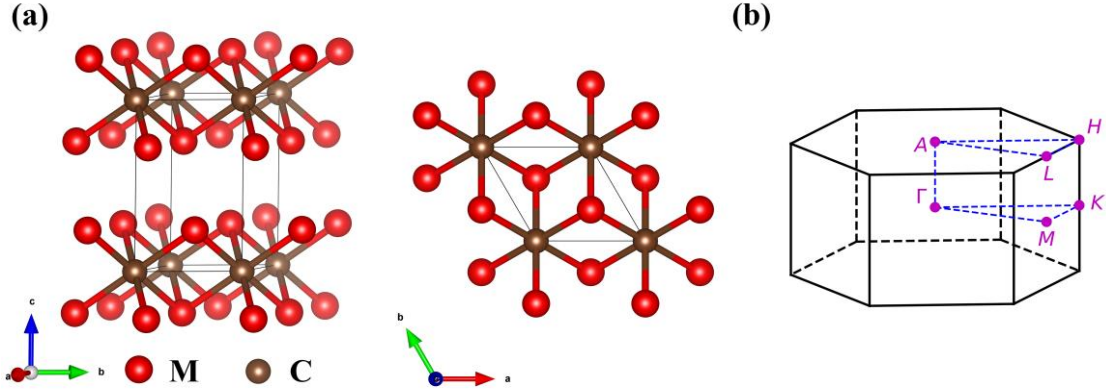

FIG. 1. (a) Crystal structure of bulk $M_2C$ ($M$=V, Nb, Ta) compounds. (b) Brillouin zone (BZ) of a primitive cell of $M_2C$.

Based on the optimized structures, the calculated band structures in the absence of SOC for Ta$_2$C are shown in Fig. 2(a), respectively, and those of other compounds are displayed in Fig. S1 in the SM [31]. The orbital character analysis shows that the Ta 5$d$ orbitals dominate the bands near the $E_F$. It is interesting that several linear bands cross each other and form the nodal points P1 and P2 and nodal line L1. Here, we use $N$ to replace the number of occupied bands, shown as the red line in Fig. 2(a). Along K-Γ and L-H route, P1 and P2 nodal points are formed by the band crossing between the singlet B and A states under the $C_2$ point group. On the other hand, it is observed that the $N$th and ($N$+1)th bands degenerate and induce the nodal line L1, which belongs to the doublet E states under $C_{3v}$ point group symmetry. The multiple band degeneracies around the $E_F$ imply the nodal-line structures in Ta$_2$C. To confirm our physical intuition, we perform a systemic nodal-line searching in the first BZ (see Methods section in the SM [31]). Here, we first identify all the *k* points with zero energy gap between the *N*th and (*N*+1)th bands and plot them with green lines in Fig. 2(b) and Fig. 2(c). Then similar treatment is used in the nodal line searching between the (*N*-1)th and *N*th bands, shown with blue lines. There are totally 17 nodal lines that can be classified into four classes: Class I, the closed nodal ring around the Γ point in the $\sigma_h$ plane; Class II, three curved nodal lines and one straight nodal line crossing the Γ point along the Γ-A route; Class III, six parabolic-like nodal lines connecting horizontal edges of the BZ; Class IV, six straight nodal lines along vertical edges of the BZ.

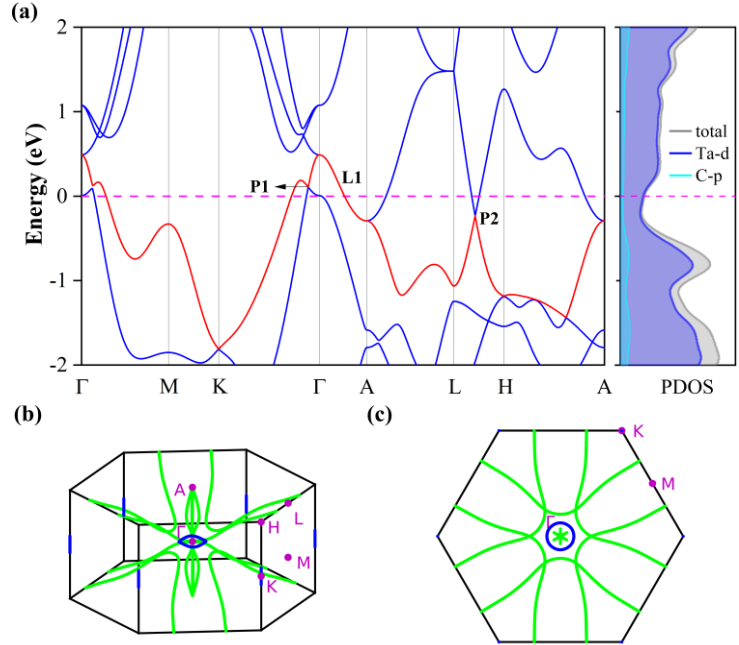

FIG. 2. (a) band structure and projected density of states of Ta$_2$C without SOC. P1, P2 and L1 represent the two nodal points and the nodal line in the band structure, respectively. (b) Band-crossing-formed nodal lines in the first BZ of Ta$_2$C. (c) top view of (b). The $E_F$ is set to zero in (a).

When the SOC effect is included, the spin-rotation symmetry is destroyed, which breaks the nodal line structure protected by *PT* symmetry [32,33]. In Fig. 3(a), the SOC lifts the degeneracies of most nodal points in band structures of Ta$_2$C. The SOC-induced bandgap is 25 meV and 360 meV at P1 and P2, and a maximum value of 550 meV at Γ point along Γ-A route, which is significantly larger than V$_2$C (from 8 to 47 meV) and Nb$_2$C (from 11 to 142 meV) (see detailed band structure of *M*$_2$C systems in Fig. S1 in the SM [31]). This means Ta has a much stronger SOC than V and Nb. Since strong SOC-induced band anticrossings around the nodal lines can generate large local SBC, it is expected that $M_2$C systems have strong SHE. To evaluate the SHE in $M_2$C, we have estimated their SHC values at $E_F$ and listed them in Table I (see SM [31] for computational details). At first glance, three SHC components are near isotropic for $V_2$C, $Nb_2$C, and $Ta_2$C. Taking $\sigma_{xy}^z$ as an example, the magnitude of SHC values increases rapidly from top to bottom, which is in accord with the variation of SOC strength from V to Ta. The SHC at $E_F$ reaches -1084 $(\hbar/e)(\Omega\ cm)^{-1}$ for $Ta_2$C, which is comparable to 5d transition metal $\beta$–W ~ 1255 $(\hbar/e)(\Omega\ cm)^{-1}$ [34], and recently reported nodal-line systems, such as InBi ~ 1100 $(\hbar/e)(\Omega\ cm)^{-1}$ [17], HfH$_2$ ~ 1100 $(\hbar/e)(\Omega\ cm)^{-1}$ [35] and Ta$_3$Bi ~ 1700 $(\hbar/e)(\Omega\ cm)^{-1}$ [15]. We also show energy-dependent SHC ($\sigma_{xy}^z$) of $Ta_2$C in Fig. 3(c). Interestingly, for a wider range of $E_F$ shifting, the magnitude of SHC values for $Ta_2$C can still keep larger than $10^3$ $(\hbar/e)(\Omega\ cm)^{-1}$. Based on the definition of SHA described in SM [31], we estimate the $G_C$ of $M_2$C and calculate the SHA. Due to their rather small $G_C$ values, the intrinsic SHA of $Ta_2$C can reach 0.22, which is larger than that of Pt (0.068) [36] and comparable to that of $\beta$–Ta (0.15) [37].

TABLE I. Intrinsic SHC for three independent tensor elements for bulk $M_2$C ($M$=V, Nb, and Ta). The unit of SHC and $G_C$ is $(\hbar/e)(\Omega\ cm)^{-1}$ and $(\Omega\ cm)^{-1}$.

|  | $\sigma_{yz}^x$ | $\sigma_{zx}^y$ | $\sigma_{xy}^z$ | $G_C$ | SHA |
|---|---|---|---|---|---|
| V$_2$C | -408 | -369 | -348 | 1.05×10$^4$ | 0.06 |
| Nb$_2$C | -581 | -569 | -565 | 1.61×10$^4$ | 0.07 |
| Ta$_2$C | -1093 | -1167 | -1084 | 9.86×10$^3$ | 0.22 |

To further understand the origin of large SHC in Ta$_2$C, we calculate the band-projected SBC and **k**-resolved SBC at $E_F$ in Fig. 3(a) and 3(b), respectively. It is observed that the large peaks of SBC at $E_F$ mainly appear at gapped nodal lines around P1, P2 and L1 in Fig. 2(a). In addition, we also calculate the **k**-resolved SBC in (001) plane at $E_F$ in Fig. 3(d). The SBC locates around the center ring and outer Hexagon region, corresponding to nodal lines near Γ point in Fig. 2(c).

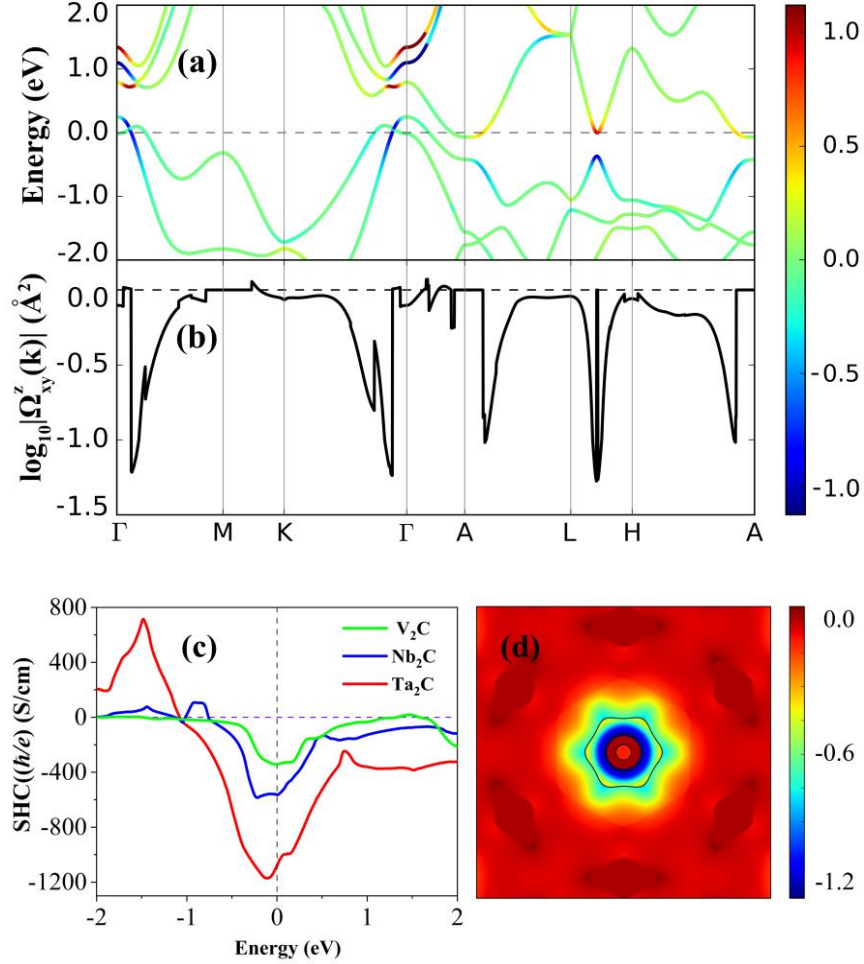

FIG. 3. (a) The band structure of Ta$_2$C with the projection of SBC. (b) **k**-revolved SBC along high symmetry lines. (c) Energy-dependent SHC of $M_2$C. (d) SBC in (001) planes of BZ. The $E_F$ is set to zero in both of (a) and (c).

## B. The electronic band structures and SHE in few-layer $M_2$C

The intimate relationship between SHC with nodal-line structure in bulk $M_2$C implies that strong SHE can also exist in monolayer or few-layered $M_2$C. We first calculate the SHC of the monolayered (1L) $M_2$C, shown in Table II. Due to the reduced symmetry of 2D structures, the dominant independent component is $\sigma_{xy}^z$. The SHC ($\sigma_{xy}^z$) increases gradually as SOC strength becomes larger from V$_2$C to Nb$_2$C and Ta$_2$C. The SHC and SHA of Ta$_2$C reach -191 $(\hbar/e)(\Omega\ \text{cm})^{-1}$ and 0.05 at $E_F$. Since V$_2$C, Nb$_2$C, and Ta$_2$C shares similar geometry and band structures, therefore, we take Ta$_2$C as an example to study the layer dependence of SHC ($\sigma_{xy}^z$) by enhancing the thickness from monolayer (1L) up to eight layers (8L). The absolute value of SHC at $E_F$ versus the number of layers is calculated and shown in Fig. 4. Generally, the SHC shows a linearly increasing trend as layer numbers go from 1 to 5 and rise slightly for 6L-8L $Ta_2C$. The maximum value of SHC is 590 $(\hbar/e)(\Omega\ cm)^{-1}$, which is larger than 2D systems such as 2L $PtSe_2$ (~27 $(\hbar/e)(\Omega\ cm)^{-1}$ at $E_F$ [24]), 3L $MoTe_2$ (~200 $(\hbar/e)(\Omega\ cm)^{-1}$ at $E_F$ [21]), and 1L SnTe (~245 $(\hbar/e)(\Omega\ cm)^{-1}$ at $E_F$ + 1.09 eV [38]).

TABLE II. Intrinsic SHC for three tensor elements for monolayer $M_2C$ ($M$=V, Nb, and Ta). The unit of SHC and $G_C$ is $(\hbar/e)(\Omega\ cm)^{-1}$ and $(\Omega\ cm)^{-1}$.

|  | $\sigma_{xy}^z$ | $\sigma_{yz}^x$ | $\sigma_{zx}^y$ | $G_C$ | SHA |
|---|---|---|---|---|---|
| $V_2C$ | -24 | -3 | -3 | 5.32×10³ | 0.009 |
| $Nb_2C$ | -46 | -6 | -6 | 1.19×10⁴ | 0.007 |
| $Ta_2C$ | -191 | -15 | -14 | 7.06×10³ | 0.05 |

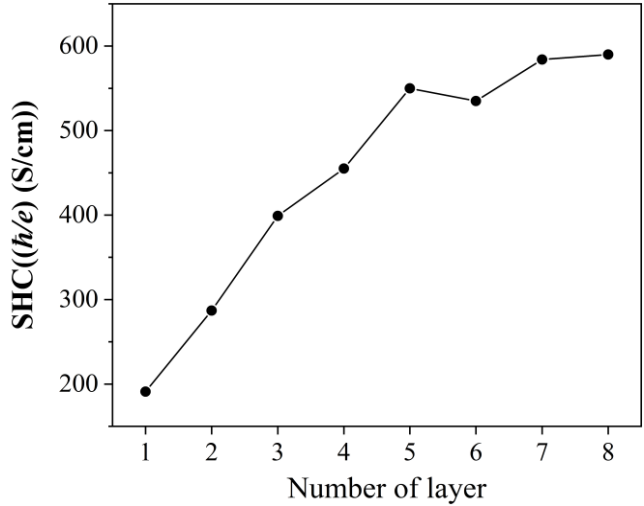

FIG. 4. Layer-dependent SHC for 1L-8L $Ta_2C$.

To figure out the origin of layer-dependent SHE observed in few-layered $Ta_2C$, we first discuss the electronic structures for 1L $Ta_2C$ in Fig. 5(a). Without SOC, the valence band (marked with a red line) crosses the conduction band at two points along the K-Γ route. In Fig. 5(e), a careful examination of the energy difference between the two bands reveals that the nodal point near the Γ point belongs to the unclosed nodal ring around the Γ point. The nodal point near the K point is one of the six equivalent points each on one Γ-K (K') route due to the symmetry protections of $C_{3v}$. In Fig. S2, the inclusion of SOC lifts the degenerate points and opens remarkable band gaps around the $E_F$ along the K-Γ route. The coexistence of nodal lines and large SOC-induced band gaps aligned near the $E_F$ induces large SBC in 1L Ta$_2$C. In Fig. 5(i), the dominant amplitude of SBC mainly concentrates around the central ring and outer raindrop-shaped area. For 3L and 5L Ta$_2$C, the increase in thickness induces repeated crossings between the two bands on the Γ-K route. Meanwhile, the conduction band sticks closely to the valence band along the M-K route for 5L Ta$_2$C. The increased degeneracy forms another circle around the Γ point and many more nodal points along the Γ-K route. Correspondingly, the inclusion of SOC leads to band gaps around the nodal lines, thus generating denser distribution of SBC and larger SHC at $E_F$. It should be noted that the energy level of the nodal points along the M-K route is not in the vicinity of $E_F$ and thus does not contribute to the SBC distribution in Fig. 5(k). In addition, for 7L Ta$_2$C, the band crossing along the M-K route induce more nodal points in the Γ-K route, which, however, do not bring much broader SBC distribution near the $E_F$, thus leading to the slow increment of SHC.

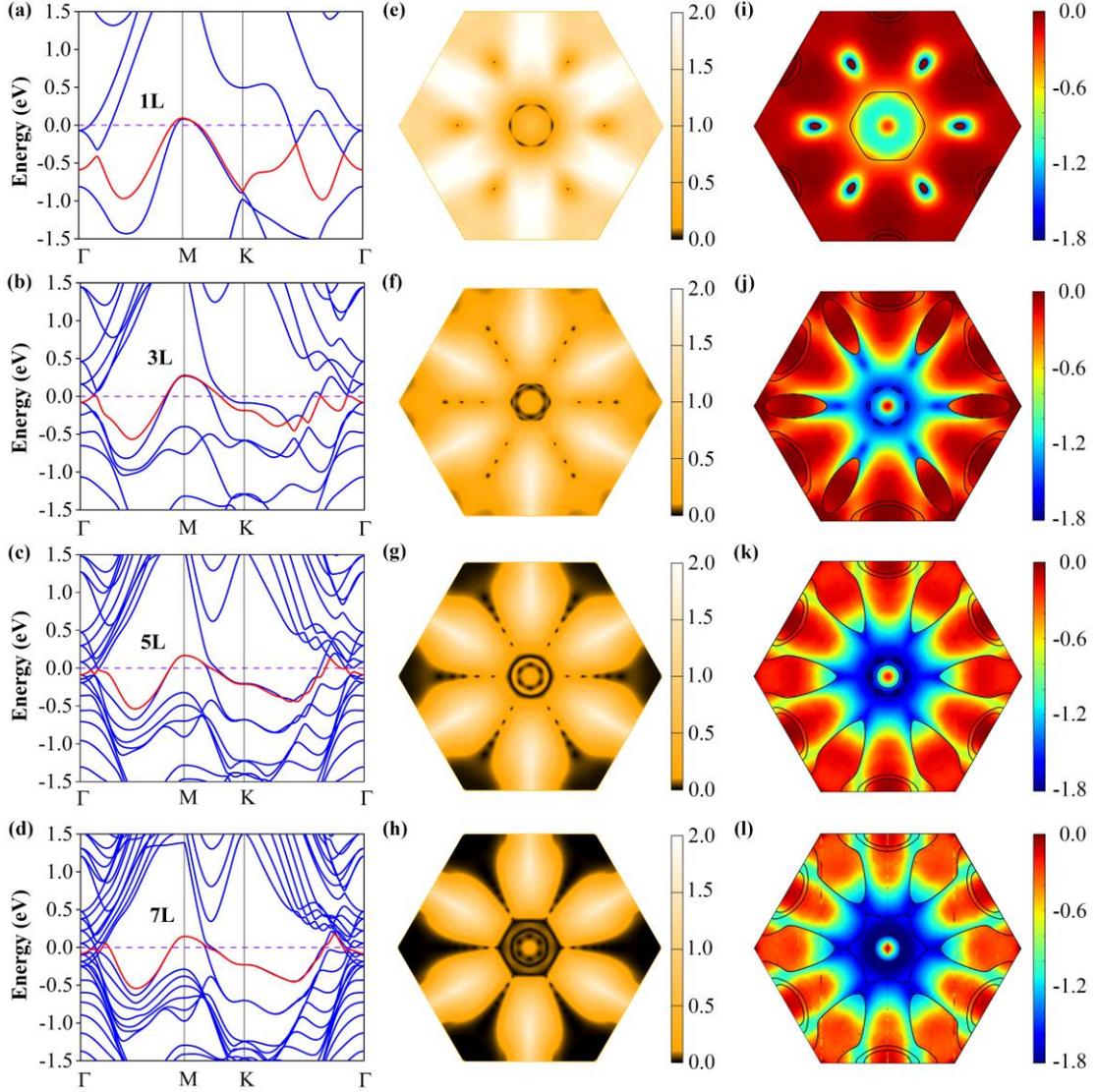

FIG. 5. (a)-(d) Electronic band structure without SOC for 1L, 3L, 5L, and 7L $Ta_2C$, respectively. (e)-(h) 2D contour of the energy gap difference between the valence and conduction bands corresponding to (a)-(d). (i)-(l) Calculated SBC in the first BZ corresponding to (a)-(d).

## III. DISCUSSION AND CONCLUSIONS

We take $M_2C$ (*M*=V, Nb, and Ta) as a representative in the above presentation. As shown in Fig. S3, the essential band crossing between the bands near the $E_F$ is also shared by other members of MXenes family. In most materials, like $Zr_2C$, $Hf_2C$, $Nb_2N$, and $Ta_2N$, the band crossing points are below or above the $E_F$, while in other materials such as $Zr_2N$ and $Hf_2N$, the band crossing points are just at the $E_F$, which are expected to contribute to strong SHC peaks.

In summary, we have predicted large intrinsic SHE in layered transition metal carbides $M_2C$ ($M$=V, Nb, and Ta). Due to strong SOC and contributions of multiple nodal lines in the band structure, the SHC and SHA of bulk $Ta_2C$ can reach up to ~1100 $(\hbar/e)(\Omega\ cm)^{-1}$ and ~0.20, respectively. For a few-layered $Ta_2C$, the $\sigma_{xy}^z$ component of SHC exhibits a monotonic increase as the number of layers increases. The maximum value of SHC is 590 $(\hbar/e)(\Omega\ cm)^{-1}$ for 8L $Ta_2C$, which is larger than reported 2D systems so far and comparable to many 3D topological systems. Therefore, one can effectively tune the SHC by controlling the layer thickness in layered 2D materials. Our results not only elucidate the interplay between layer-dependent SHC and band topology, but also provide theoretical guidance for developing next-generation spintronic devices.

## ACKNOWLEDGMENTS

This work is supported by National Natural Science Foundation of China (Grant No. 11574118), NSFC (Grants No. 51672023), and NSAF (Grant No. U1930402).

## References

[1] J. E. Hirsch, Spin Hall Effect, Phys. Rev. Lett. **83**, 1834 (1999).

[2] J. Sinova, S. O. Valenzuela, J. Wunderlich, C. H. Back, and T. Jungwirth, Spin Hall Effects, Rev. Mod. Phys. **87**, 1213 (2015).

[3] G. Y. Guo, S. Murakami, T.-W. Chen, and N. Nagaosa, Intrinsic Spin Hall Effect in Platinum: First-Principles Calculations, Phys. Rev. Lett. **100**, 096401 (2008).

[4] J. Qiao, J. Zhou, Z. Yuan, and W. Zhao, Calculation of Intrinsic Spin Hall Conductivity by Wannier Interpolation, Phys. Rev. B **98**, 214402 (2018).

[5] N. H. D. Khang, Y. Ueda, and P. N. Hai, A Conductive Topological Insulator with Large Spin Hall Effect for Ultralow Power Spin–Orbit Torque Switching, Nat. Mater. **17**, 808 (2018).

[6] D.-F. Shao, G. Gurung, S.-H. Zhang, and E. Y. Tsymbal, Dirac Nodal Line Metal for Topological Antiferromagnetic Spintronics, Phys. Rev. Lett. **122**, 077203 (2019).

[7] K. Kondou, H. Chen, T. Tomita, M. Ikhlas, T. Higo, A. H. MacDonald, S. Nakatsuji, and Y. Otani, Giant Field-like Torque by the out-of-Plane Magnetic Spin Hall Effect in a Topological Antiferromagnet, Nat Commun **12**, 6491 (2021).

[8] Y. Fan et al., Magnetization Switching through Giant Spin-Orbit Torque in a Magnetically Doped Topological Insulator Heterostructure, Nat. Mater. **13**, 699 (2014).

[9] Y. Wang, P. Deorani, K. Banerjee, N. Koirala, M. Brahlek, S. Oh, and H. Yang, Topological Surface States Originated Spin-Orbit Torques in $Bi_2Se_3$, Phys. Rev. Lett. **114**, 257202 (2015).

[10] Y. Yen and G.-Y. Guo, Tunable Large Spin Hall and Spin Nernst Effects in the Dirac Semimetals ZrXY ( X = Si , Ge ; Y = S , Se , Te), Phys. Rev. B **101**, 064430 (2020).

[11] Y. Sun, Y. Zhang, C. Felser, and B. Yan, Strong Intrinsic Spin Hall Effect in the TaAs Family of Weyl Semimetals, Phys. Rev. Lett. **117**, 146403 (2016).

[12] T. Ng, Y. Luo, J. Yuan, Y. Wu, H. Yang, and L. Shen, Origin and Enhancement of the Spin Hall Angle in the Weyl Semimetals LaAlSi and LaAlGe, Phys. Rev. B **104**, 014412 (2021).

[13] Y. Sun, Y. Zhang, C.-X. Liu, C. Felser, and B. Yan, Dirac Nodal Lines and Induced Spin Hall Effect in Metallic Rutile Oxides, Phys. Rev. B **95**, 235104 (2017).

[14] E. Derunova, Y. Sun, C. Felser, S. S. P. Parkin, B. Yan, and M. N. Ali, Giant Intrinsic Spin Hall Effect in $W_3$Ta and Other A15 Superconductors, Sci. Adv. **5**, eaav8575 (2019).

[15] W. Hou, J. Liu, X. Zuo, J. Xu, X. Zhang, D. Liu, M. Zhao, Z.-G. Zhu, H.-G. Luo, and W. Zhao, Prediction of Crossing Nodal-Lines and Large Intrinsic Spin Hall Conductivity in Topological Dirac Semimetal $Ta_3$As Family, Npj Comput Mater **7**, 37 (2021).

[16] N.-N. Zhao, K. Liu, and Z.-Y. Lu, Large Intrinsic Spin Hall Conductivity and Spin Hall Angle in the Nodal-Line Semimetals $ThAl_2$ and $ThGa_2$, Phys. Rev. B **105**, 235119 (2022).

[17] Y. Zhang, Q. Xu, K. Koepernik, C. Fu, J. Gooth, J. van den Brink, C. Felser, and Y. Sun, Spin Nernst Effect in a P-Band Semimetal InBi, New J. Phys. **22**, 093003 (2020).

[18] G. M. Stiehl et al., Layer-Dependent Spin-Orbit Torques Generated by the Centrosymmetric Transition Metal Dichalcogenide $\beta-MoTe_2$, Phys. Rev. B **100**, 184402 (2019).

[19] Q. Shao, G. Yu, Y.-W. Lan, Y. Shi, M.-Y. Li, C. Zheng, X. Zhu, L.-J. Li, P. K. Amiri, and K. L. Wang, Strong Rashba-Edelstein Effect-Induced Spin–Orbit Torques in Monolayer Transition Metal Dichalcogenide/Ferromagnet Bilayers, Nano Lett. **16**, 7514 (2016).

[20] G. M. Stiehl, D. MacNeill, N. Sivadas, I. El Baggari, M. H. D. Guimarães, N. D. Reynolds, L. F. Kourkoutis, C. J. Fennie, R. A. Buhrman, and D. C. Ralph, Current-Induced Torques with Dresselhaus Symmetry Due to Resistance Anisotropy in 2D Materials, ACS Nano **13**, 2599 (2019).

[21] P. Song et al., Coexistence of Large Conventional and Planar Spin Hall Effect with Long Spin Diffusion Length in a Low-Symmetry Semimetal at Room Temperature, Nat. Mater. **19**, 292 (2020).

[22] B. Zhao, D. Khokhriakov, Y. Zhang, H. Fu, B. Karpiak, A. Md. Hoque, X. Xu, Y. Jiang, B. Yan, and S. P. Dash, Observation of Charge to Spin Conversion in Weyl Semimetal $WTe_2$ at Room Temperature, Phys. Rev. Research **2**, 013286 (2020).

[23] H. Xu et al., High Spin Hall Conductivity in Large- Area Type- II Dirac Semimetal $PtTe_2$, Adv. Mater. **32**, 2000513 (2020).

[24] J. Li, H. Jin, Y. Wei, and H. Guo, Tunable Intrinsic Spin Hall Conductivity in Bilayer $PtTe_2$ by Controlling the Stacking Mode, Phys. Rev. B **103**, 125403 (2021).

[25] Y. Gogotsi and B. Anasori, The Rise of MXenes, ACS Nano **13**, 8491 (2019).

[26] M. Naguib, M. W. Barsoum, and Y. Gogotsi, Ten Years of Progress in the Synthesis and Development of MXenes, Adv. Mater. **33**, 2103393 (2021).

[27] G. Gao, G. Ding, J. Li, K. Yao, M. Wu, and M. Qian, Monolayer MXenes: Promising Half-Metals and Spin Gapless Semiconductors, Nanoscale **8**, 8986 (2016).

[28] A. Sufyan, A. B. Maghirang, G. Macam, Z.-Q. Huang, C.-H. Hsu, and F.-C. Chuang, Electronic and Topological Band Evolution of VB-Group Transitionmetal Monocarbides $M_2$C (M=V, Nb, or Ta) Bulk and Monolayer, Mater. Today Commun. **32**, 103875 (2022).

[29] Y. Luo, C. Cheng, H.-J. Chen, K. Liu, and X.-L. Zhou, Systematic Investigations of the Electron, Phonon and Elastic Properties of Monolayer $M_2$C (M = V, Nb, Ta) by First-Principles Calculations,

J. Phys.: Condens. Matter **31**, 405703 (2019).

[30] N. J. Lane, M. W. Barsoum, and J. M. Rondinelli, Correlation Effects and Spin-Orbit Interactions in Two-Dimensional Hexagonal 5d Transition Metal Carbides, $Ta_{n+1}C_n$ (n = 1,2,3), Europhys. Lett. **101**, 57004 (2013).

[31] See Supplemental Material at http://link.aps.org/supplemental/xxxx for a description of the density functional methods and additional DFT results.

[32] Y. Kim, B. J. Wieder, C. L. Kane, and A. M. Rappe, Dirac Line Nodes in Inversion-Symmetric Crystals, Phys. Rev. Lett. **115**, 036806 (2015).

[33] Y. X. Zhao, A. P. Schnyder, and Z. D. Wang, Unified Theory of PT and CP Invariant Topological Metals and Nodal Superconductors, Phys. Rev. Lett. **116**, 156402 (2016).

[34] X. Sui, C. Wang, J. Kim, J. Wang, S. H. Rhim, W. Duan, and N. Kioussis, Giant Enhancement of the Intrinsic Spin Hall Conductivity in $\beta$-Tungsten via Substitutional Doping, Phys. Rev. B **96**, 241105 (2017).

[35] X. Zuo et al., Giant and Robust Intrinsic Spin Hall Effects in Metal Dihydrides: A First-Principles Prediction, Phys. Rev. B **103**, 125159 (2021).

[36] Y. Wang, P. Deorani, X. Qiu, J. H. Kwon, and H. Yang, Determination of Intrinsic Spin Hall Angle in Pt, Appl. Phys. Lett. **105**, 152412 (2014).

[37] L. Liu, C.-F. Pai, Y. Li, H. W. Tseng, D. C. Ralph, and R. A. Buhrman, Spin-Torque Switching with the Giant Spin Hall Effect of Tantalum, Science **336**, 555 (2012).

[38] J. Sławińska, F. T. Cerasoli, H. Wang, S. Postorino, A. Supka, S. Curtarolo, M. Fornari, and M. Buongiorno Nardelli, Giant Spin Hall Effect in Two-Dimensional Monochalcogenides, 2D Mater. **6**, 025012 (2019).